\documentstyle[secnumeq]{fbssuppl}
\title{$NN$ and $N\Delta$ Form Factors viewed from ChPT}
\author{Thomas R. Hemmert\thanks{{\it E-mail address}: th.hemmert@fz-juelich.de}
\thanks{Invited Lecture at the joint ECT*/Jefferson Lab Workshop on N* physics and 
non-perturbative QCD, Trento, Italy, May 18 - 29, 1998; {\tt [FZJ-IKP(Th)-1998-17]}}
}
\institute{IKP (Th), FZ-J{\" u}lich, D-52425 J{\" u}lich}
\begin{document}
\maketitle
\begin{abstract}
I discuss recent work on nucleon form factors, magnetic strangeness in the nucleon
 and the isovector nucleon-delta transition based on the broken chiral symmetry of QCD 
utilizing recent theoretical developments in ChPT. 
\end{abstract}
\section{Introduction}
During this workshop we have heard both theoretical and experimental 
presentations looking for the onset of {\it perturbative} QCD formulated in explicit 
(current) quark and gluon degrees of freedom at moderate/high four-momentum transfer, 
{\it e.g.} $Q^2>1$ GeV$^2$. At lower momentum transfer one cannot avoid the 
complications of the strong coupling regime. In 
this talk I want to discuss some of the constraints resulting from the broken chiral 
symmetry of QCD for baryon form factors in the {\it non-perturbative} regime of QCD 
\cite{bfhm,hms,ghkp}. 

\section{Nucleon Form Factors and ChPT}

The chiral symmetry of the light flavor sector of QCD is spontaneously broken at low 
energies leading to the existence of Goldstone boson modes. Here we focus on the 
(u,d)-quark
sector only which leads to the identification of the pions as the Goldstone Bosons. All 
low energy dynamics is governed by these lightest hadronic degrees of freedom and the 
chiral symmetry puts very strict constraints on their interactions among themselves, 
with external sources and on their coupling to matter fields (baryons, etc.). This 
Goldstone-boson dominated regime of {\it non-perturbative} QCD at low energies can be 
formulated exactly in an effective lagrangian formalism called Chiral Perturbation 
Theory (ChPT) \cite{GL}. With the pions being the lightest degrees of freedom in the 
hadron spectrum, ChPT suggests that the long range structure of baryons and its leading 
momentum dependence is governed by the chiral symmetry of the pion interaction. 

The electro-weak structure of baryons is parameterized via form factors. In the case of 
the nucleon 
they have been analyzed in one-loop relativistic baryon 
ChPT \cite{GSS} and in a non-relativistic approach called HBChPT \cite{BKKM} in the 
past. Recently \cite{bfhm}, we have repeated this analysis utilizing a 
phenomenological extension of ChPT called the ``small scale expansion'' \cite{SSE}. 
In this approach one includes the first nucleon resonance $\Delta$(1232) as an explicit 
degree of freedom in a phenomenologically resummed chiral expansion. 
In \cite{bfhm} all 6 form factors of the nucleon are discussed, here I will only 
address the isovector Pauli form factor $F_2^v(q^2)$.

Consider the nucleon matrix element of the isovector component of the
quark vector current $V_\mu^i =\bar q \gamma_\mu (\tau^i /2) q$, 
which involves a vector (Dirac) and a tensor (Pauli) form factor,
\begin{eqnarray}
\langle N(p_2)|V_{\mu}^i(0)|N(p_1)\rangle=\bar{u}(p_2)\left[F_{1}^{\;v}(q^2)\;
\gamma_\mu+\frac{i}{2M_N}F_{2}^{\;v}(q^2)\;\sigma_{\mu\nu}
q^\nu\right]u(p_1) \, ,
\end{eqnarray}
where $u(p)$ is a Dirac spinor and
$q^2 = (p_2 - p_1)^2$ is the invariant momentum transfer squared. The radii of these 
form factors should be determined by the extension 
of the pion cloud. {\it E.g.} for the radius of $F_2^v(q^2)$ one finds to leading order 
in HBChPT \cite{BKKM}, [ SSE \cite{bfhm} ]
\begin{eqnarray}
\left(r_{2}^{v}\right)^2&=&\frac{g_{A}^2 M_N}{8 F_{\pi}^2 \kappa_v \pi m_\pi}
                           +\left[\frac{g_{\pi N\Delta}^2 M_N}{9 F_{\pi}^2 \kappa_v
                           \pi^2 \sqrt{\Delta^2-m_{\pi}^2}} \log\left[
                           \frac{\Delta}{m_\pi}+\sqrt{\frac{\Delta^2}{
                           m_{\pi}^2}-1}\right] \right] \nonumber \\
                        &=&0.52{\rm fm}^2 \; [+\; 0.09 {\rm fm}^2] \; , \label{eq:r2}
\end{eqnarray}
compared with the empirical value, $(r_2^v)^2 = 0.80$fm$^2$ \cite{MMD}. The only 
parameters are the pion decay constant (mass) $F_\pi\,(m_\pi)$, the $\pi NN\,(\pi\Delta 
N)$ couplings $g_A\,(g_{\pi N\Delta})$, the nucleon mass $M_N$, the mass-splitting
$\Delta=M_\Delta -M_N$ and the anomalous 
isovector magnetic moment $\kappa_v$. One can see that
already the {\it leading} HBChPT result for the extension of the pion cloud provides
a good estimate for the size of the nucleon in this channel. Inclusion of explicit delta
components in the nucleon wavefunction around which the pions can fluctuate provides 
a 17\% correction in the right direction \cite{bfhm}. In the chiral limit, we recover the
well known $1/m_\pi$ singularity, which is not touched by the resonance contribution in 
accord with general decoupling requirements. Other form factor results are discussed in 
\cite{bfhm}.

\section{Magnetic Strangeness in the Nucleon}

So far we have focused on baryon ChPT involving two light flavors. The analysis 
presented in the previous section can be generalized in a straightforward fashion to a 
SU(3) chiral symmetry of QCD, {\it i.e.} the inclusion of explicit strange degrees of 
freedom like kaons, lambdas etc. Repeating the analysis of the isovector form factors 
of the nucleon in SU(3) HBChPT one finds an extra contribution from kaon loops to 
$(r_2^v)^2$ (Eq.(\ref{eq:r2})) of the order of a few percent. This contribution vanishes 
if the strange quark mass becomes very heavy. For the case of the isovector nucleon 
form factors SU(3) HBChPT therefore 
reproduces our physical expectation that the kaon contributions are much less important 
than their pion counterparts due to the much larger mass. Here the pions clearly 
dominate the long-range physics and control the size of the nucleon. This is also 
seen in the spectral function \cite{spec}.

However, this is not always the case. For example in the isoscalar form factors of the 
nucleon the leading chiral contribution to the radius is given by the kaon cloud, as 
the pionic contribution only begins via 3 pion intermediate states at the 2 loop level. 
The ${\cal O}(p^3)$ analysis suggests that roughly 30\% of the isoscalar radius of the 
nucleon comes from structure related to the kaons in the nucleon! For details regarding 
these issues we refer to \cite{bfhm,hms,spec}.  

Another sector where explicit strange degrees of freedom figure prominently 
concerns--quasi by definition--the recent interest in the so-called ``strangeness 
content of the nucleon'', {\it e.g.} see ref.\cite{report}. In the following we focus 
on the 
strangeness vector current of the nucleon defined as
\begin{equation}\label{svc}
\langle N|\;\bar{s}\;\gamma_\mu\; s\;|N \rangle
= \langle N|\;\bar{q}\;\gamma_\mu\;
(\lambda^0/3-\lambda^8/\sqrt{3}) \; q\;| N \rangle 
                                  = (1/3)J_{\mu}^0- (1/\sqrt{3})J_{\mu}^8 \; ,
\end{equation}
with $q=(u,d,s)$ denoting the triplet of the light quark fields and
$\lambda^0 = I\; (\lambda^a)$ 
the unit (the $a=8$ Gell--Mann) SU(3) matrix.
Assuming conservation of  all vector currents, the corresponding singlet and octet 
vector current for a nucleon can then be written as 
\begin{equation}\label{curr}
J_{\mu}^{0,8}
=\bar{u}_N(p^\prime)\left[F_{1}^{(0,8)}(q^2)\gamma_\mu+ F_{2}^{(0,8)}(q^2)
        \frac{i\sigma_{\mu\nu}q^\nu}{2M_N}\right] u_N(p) \; .
\end{equation}
Here, $q_\mu=p^{\prime}_\mu-p_\mu$ corresponds to the four--momentum transfer to the 
nucleon by the external singlet ($v_{\mu}^{(0)}=v_\mu \lambda^0$) and the octet ($
v_{\mu}^{(8)}=v_\mu\lambda^8$) vector source $v_\mu$, respectively.
The strangeness Dirac and Pauli form factors are defined via
\begin{equation}
F_{1,2}^{(s)}(q^2)= \frac{1}{3}F_{1,2}^{(0)}(q^2)
-\frac{1}{\sqrt{3}}F_{1,2}^{(8)}(q^2) \; ,
\end{equation}
subject to the normalization $F_1^{(s)} (0) = 0, F_{2}^{(s)} (0) =\kappa_{B}^{(s)}$, with
$\kappa_{B}^{(s)}$ the  (anomalous) strangeness moment. 
In the following we concentrate our analysis on the ``magnetic'' strangeness
form factor $G_{M}^{(s)}(q^2)$, which in analogy to the (electro)magnetic
Sachs form factor is defined as
\begin{eqnarray}
G_{M}^{(s)}(q^2)=F_{1}^{(s)}(q^2)+F_{2}^{(s)}(q^2)~. \label{eq:def1}
\end{eqnarray}
In the case of a nucleon $G_M^{(s)}(0)\equiv \mu_N^{(s)}$ defines the so 
called ``strange magnetic moment'' of the nucleon whose sign/size is heavily contested
in theoretical analyses. Furthermore, it is precisely this form factor at 
$q^2=-0.1$GeV$^2$ 
which has been analyzed in the recent Bates measurement~\cite{SAMPLE}.

One expects that for low $q^2$ ChPT can give a prediction for this 
(as of 1998!) unknown quantity. For the case of $\mu_N^{(s)}$ this is only partially 
correct as one needs additional information about an unknown isosinglet counterterm 
\cite{hms,im}. However, even if one cannot calculate the overall normalization of 
$G_M^{(s)}(q^2)$ at $q^2=0$, the evolution of this form factor with $q^2$ can be 
predicted in terms of well-known low energy quantities! To ${\cal O}(p^3)$ in SU(3) 
HBChPT one finds \cite{hms}
\begin{eqnarray}\label{eq:c}
G_{M}^{(s)}(Q^2) &=& \mu_{N}^{(s)}+\frac{\pi M_N m_K}{(4\pi
F_{\pi})^2}\;\frac{2}{3}\left( 5 D^2-6 D F+9 F^2 \right) \, f(Q^2)~,
\label{eq:gms}
\end{eqnarray}
with $Q^2=-q^2,D\simeq 3/4,F\simeq 1/2,F_\pi \equiv  (F_\pi + F_K)/2 \simeq 102\,$MeV 
the average pseudoscalar decay constant and $m_K$ being the kaon mass.  
The momentum dependence is given entirely in terms of the function
\begin{equation}\label{f2q}
f(Q^2) = -\frac{1}{2} + \frac{4+Q^2/m_K^2}{4\sqrt{Q^2/m_K^2}} 
\arctan \biggl( \frac{\sqrt{Q^2}}{2m_K} \biggr)~. 
\end{equation}
For small and moderate $Q^2$ it rises almost linearly with increasing $Q^2$.

I emphasize that Eq.(\ref{eq:gms}) only contains the leading order chiral contribution 
which stems exclusively from the kaon-cloud of the nucleon. It will be interesting to 
calculate the next-to-leading order ({\it i.e.} ${\cal O}(p^4)$) correction to this 
result in order to check possible contributions from vector mesons which are usually 
assumed to dominate this form factor \cite{vec}. However, already at ${\cal O}(p^3)$ 
one can implicitly include some of the higher order corrections if one analyzes the 
magnetic isoscalar form factor $G_M^{I=0}(Q^2)$ and the strange magnetic form factor 
$G_M^{(s)}(Q^2)$ simultaneously \cite{hms}. One obtains the {\it model-independent} 
connection
\begin{eqnarray}
G_{M}^{(s)}(Q^2) = \mu_{N}^{(s)}+\mu_N^{I=0}-G_{M}^{I=0}(Q^2)+{\cal O}(p^4) \; ,
\label{eq:mi}
\end{eqnarray}
where $\mu_N^{I=0}$ denotes the isoscalar magnetic moment of the nucleon. To 
${\cal O}(p^3)$ 
one therefore predicts that the low $Q^2$ behavior of the strange magnetic form factor 
of the nucleon is {\it exactly} controlled by the well-known isoscalar form factor of 
the nucleon! For details I refer to \cite{hms}. 

Eqs.(\ref{eq:gms},\ref{eq:mi}) can be considered as a lower, upper bound on the $q^2$ 
evolution of the strange magnetic form factor at low momentum transfer \cite{hms}. 
Both relations 
can be used to extrapolate from the experimentally determined values for 
$G_M^{(s)}(Q^2)$ at $Q^2>0$ to the sought after strange magnetic 
moment $\mu_N^{(s)}$ of the nucleon at $Q^2=0$. Clearly, with improving experimental 
accuracy on $G_M^{(s)}(Q^2)$ one also needs to calculate the ${\cal O}(p^4)$ 
corrections to both relations. Furthermore, comparing Eqs.(\ref{eq:gms},\ref{eq:mi}) we 
are also looking forward to the mapping 
of the low $Q^2$ dependence of $G_M^{(s)}(Q^2)$ by the G0 collaboration at 
J-Lab \cite{G0}.

\section{The Isovector $N\Delta$ Transition}

Finally, I want to give a brief update on the ongoing calculations \cite{ghkp,photo} 
regarding the isovector nucleon-delta transition multipoles and form factors. Recent 
interest is mainly triggered by three observations:
\begin{enumerate}
\item 
In a multipole analysis one finds that in the photoexcitation of $\Delta$(1232) 
[$\gamma N\rightarrow \Delta$] one can only have magnetic dipole (M1) or electric 
quadrupole (E2) transitions from the nucleon to the delta. Simple constituent quark
 models of the nucleon however generally assume all quarks to be in an s-wave state and 
therefore predict zero strength for the E2 transition. Several fits to pion 
photoproduction data in the delta region however show a non-zero ratio of E2/M1 strength
 of about -1\% to -3\% ({\it e.g.} \cite{e2m1}), indicating non-radial/many-body 
components in the ground-state wave function of the nucleon.

\item
For electroproduction of $\Delta$(1232) [$\gamma^\ast N\rightarrow\Delta$] 
the transition multipoles M1,E2 do not only develop a dependence on the four-momentum 
transfer (squared) $Q^2$ but one can now have additional contributions from a Coulomb 
quadrupole transition C2. Our knowledge of the $Q^2$ dependence of these three 
multipoles for $0<Q^2<1$GeV$^2$ mainly stems from experiments of the 1970s \cite{70s}. 
Recently, new measurements have started at Bonn which seem to validate the old analyses 
showing interesting differences in the $Q^2$ behavior among these multipoles for 
$Q^2<0.3$GeV$^2$ \cite{kal}. Furthermore, one would also like to compare the 
$Q^2$-falloff of the $N\Delta$ transition form factors with the well-known dipole 
behavior of the electric/magnetic Sachs form factors of the nucleon, {\it e.g.} see 
\cite{70s,stoler}.

\item
Perturbative QCD predicts that for very large four-momentum transfer the ratio of E2/M1 
for the case of delta electroproduction should tend to unity. At which finite $Q^2$ the 
crossover from a negative to a positive ratio should happen and whether this point is 
kinematically accessible at present/future electron scattering machines is an issue of 
current theoretical debate, {\it e.g.} \cite{CM}.
\end{enumerate}

We have started two collaborations \cite{ghkp,photo} to look into these topics from the 
viewpoint of ChPT. In particular, we are using the recently 
developed SSE formalism \cite{SSE} in order to treat the delta 
resonance in a systematic fashion. What kind of results can one expect from these 
efforts?
\begin{enumerate}
\item 
There exist already 2 calculations regarding the ratio of E2/M1 at the real photon point
 utilizing ChPT \cite{butler}. Our present understanding is that 
one needs to take into account non-zero contributions from {\it three} different 
ingredients---namely pion loops, 1/M corrections and counterterms. While the loop 
contribution is relatively easy to calculate and agreed upon, the contributions from 
the 1/M corrections and counterterms have not been handled with the same accuracy so 
far. SSE offers a systematic formalism to address both aspects. At this point we 
can say that the actual number for E2/M1 in ChPT is quite sensitive to the 
treatment/size of several unknown counterterms. In order to settle this issue one needs 
a full calculation of pion-photoproduction in the delta resonance region  
\cite{photo} in order to fix these unknowns with the accuracy required 
for E2/M1. Only then one can expect a new systematic prediction for 
E2/M1 from ChPT. We also note that in the past only the leading delta contribution to 
the s-wave multipole $E_{0+}$ had been calculated explicitly in SSE \cite{SSE}. The 
p-wave multipoles are known to receive large contributions from 
$\Delta$(1232), but so far these effects have only been included via ``resonance 
saturation'' in higher order couplings \cite{ZPC}. Utilizing SSE  
\cite{SSE}, we are now analyzing explicit $\Delta$(1232) components in the three 
p-wave multipoles. It will be interesting to see how far in energy the inclusion of 
explicit delta degrees of freedom can extend the applicability of ChPT to 
pion-photoproduction off nucleons into the delta resonance region. \cite{photo}. 

\item
Surprisingly, the determination of the $Q^2$-evolution of the three $N\Delta$ 
transition form factors and of the corresponding three transition multipoles is a much 
simpler problem in ChPT, but has not been addressed so far. The important point to 
realize is that most of the unknown couplings/counterterms only concern the $Q^2=0$ 
values. Once one fixes the form factors/multipoles at the measured real photon values 
\cite{e2m1} one obtains their $Q^2$-dependence in terms of very few parameters which 
are under control. It is then straightforward to extract radii for the transition form 
factors and compare with the form factors of the nucleon. This project \cite{ghkp} is 
close to being finished once the problem of the scaling in the radii (discussed below) 
is fixed.

\item
Concerning the third issue, ChPT can certainly not answer the problem of the onset of 
perturbative QCD in the E2/M1 ratio, probably even the zero-crossing point is at too 
high a momentum transfer for this approach. However, it should be possible to say 
whether E2/M1 first drops even more negative for low momentum transfer and whether 
there is a {\it turning point} after which the curve moves towards a positive value.
\end{enumerate}

Finally, I want to address a problem that we encountered during the calculation 
of the isovector $N\Delta$ transition form factors. 
 Assuming conservation of the vector current as well as invariance under P,C,T symmetry 
operations one concludes that in general there exist 3 independent structures for such a 
transition. To be more specific, let's assume that we are talking about the process 
$\Delta\rightarrow N\gamma^*$. The matrix-element is then typically written as 
\cite{scadron}\footnote{Minor differences to the form of Eq.(\ref{eq1}) arise via 
field-redefinitions utilizing the equations of motion for the baryons. However, this 
does not change the thrust of the above argument.} 
\begin{eqnarray}\label{eq1}
i{\cal M}_{\Delta\rightarrow N\gamma}^{full}
&=&{e\over 2 M_N} \bar{u}(p_N) \gamma_5\left[
   g_1(q^2)( \not{q} \epsilon_\mu - \not{\epsilon} q_\mu )
 + {g_2(q^2)\over 2M_N} (p_N\cdot\epsilon\, q_\mu \right. \nonumber \\
& &\left. - p_N\cdot q\, \epsilon_\mu)
 + {g_3(q^2)\over 2 M_N} (q\cdot\epsilon\, q_\mu - q^2 \epsilon_\mu)\right]
u^\mu_\Delta(p_\Delta) \ .
\end{eqnarray}
Here $M_N$ is the nucleon mass, $p_{N,\Delta}$ denotes the momentum of the nucleon, delta
and $q$, $\epsilon$ are the photon momentum
and polarization vectors, respectively. The delta is described in the
Rarita-Schwinger formalism, i.e. as an axial-vector spinor $u^\mu_\Delta$. Now one 
proceeds to calculate this matrix element in a {\it non-relativistic microscopic 
approach}, in our case SSE \cite{SSE}. Calculating to third 
order in the expansion scheme the calculation can produce up to two inverse 
powers of the expansion scale, {\it i.e.} one is sensitive to structures up to $1/M_N^2$.
 In order to match the calculation with the most general matrix element Eq.(\ref{eq1}) 
one also needs to expand it up to the same power in $1/M_N^2$. One finds
\begin{eqnarray}\label{exp}
i{\cal M}_{\Delta->N\gamma}^{(3)}& =&e \ \bar{u}_v(r_N)\left\{ 
 (S\cdot \epsilon) q_\mu   
\left[ \frac{g_1(q^2)}{M_N}+{\cal O}(1/M_N^3)  \right] \right.
\nonumber \\ 
& &+ (S\cdot q) \epsilon_\mu   
\left[ -\frac{g_1(q^2)}{M_N}-\frac{\Delta}{2M_N^2}g_1(0)+\frac{\Delta}{4M_N^2}g_2(0)
+{\cal O}(1/M_N^3) \right] 
\nonumber \\ 
& &+ (S\cdot q) (v\cdot\epsilon) q_\mu  
\left[ \frac{( g_1(0) - {1\over 2} g_2(0) )}{2M_N^2} +{\cal O}(1/M_N^3) \right] 
\nonumber \\
& &\left. + (S\cdot q) (q \cdot \epsilon) q_\mu  
\left[ 0+ {\cal O}(1/M_N^3) \right] 
\right\} u_{v,\Delta}^{\mu}(0)\ .
\end{eqnarray}
Here $S_\mu$ denotes the Pauli-Lubanski vector, $v_\mu$ corresponds to the velocity 
vector of the delta reference frame and $\Delta=M_\Delta-M_N$. As one can see from 
Eq.(\ref{exp}) the $1/M_N^2$-expansion demands that there are no explicit structures 
proportional to $\epsilon \cdot q$ to this order. Nevertheless the SSE calculations 
yield such terms! We therefore have to conclude that the often-used form for the 
isovector $N\Delta$ transition Eq.(\ref{eq1}) is {\it not compatible} with 
(non-relativistic) microscopic calculations of this transition that rely on a systematic
 1/M expansion. It is therefore mandatory to rescale $g_2(q^2), [g_3(q^2)]$ by 
$M_N/\Delta, [M_N^2/\Delta^2]$ in order to achieve a systematic matching between the 
microscopic calculations and the most general amplitude. Furthermore, without this 
rescaling of the form factors their transition radii would scale as 
$r_i^2 \sim M_N^n\; ;i=2,3;n\geq1$, {\it i.e.} one would see no 1/M 
suppression\footnote{The only known case of such an anomalous scaling concerns the 
pseudoscalar form factor of the nucleon which contains the (light) pion-pole, see 
{\it e.g.} \cite{bfhm}.} in the 
radii compared to the $q^2=0$ point! Finally, we note that phenomenological analyses of 
data utilizing Eq.(\ref{eq1}) are not affected by this problem, as long as all 
amplitudes are treated in a fully relativistic form. 
A detailed publication describing all these aspects is in preparation \cite{ghkp}.

\begin{acknowledge}
I would like to thank my collaborators V. Bernard, H.W. Fearing, G.C. Gellas, 
C.N. Ktorides, U.-G. Mei{\ss}ner, G.I. Poulis and S. Steininger and acknowledge helpful
financial support from ECT*. 
\end{acknowledge}

\end{document}